Growth rates of modern science: A latent piecewise growth curve approach to model publication numbers from established and new literature databases

Lutz Bornmann, Robin Haunschild, Rüdiger Mutz

*Science Policy and Strategy Department

Administrative Headquarters of the Max Planck Society

Hofgartenstr. 8,

80539 Munich, Germany.

E-mail: bornmann@gv.mpg.de

ORCID: 0000-0003-0810-7091

** Max Planck Institute for Solid State Research

Heisenbergstraße 1,

70569 Stuttgart, Germany.

Email: r.haunschild@fkf.mpg.de

ORCID: 0000-0001-7025-7256

*** Center for Higher Education and Science Studies, CHESS

University of Zurich

Andreasstrasse 15,

8050 Zurich, Switzerland.

Email: ruediger.mutz@uzh.ch

ORCID: 0000-0003-3345-6090

**Abstract**

Growth of science is a prevalent issue in science of science studies. In recent years, two new bibliographic databases have been introduced which can be used to study growth processes in science from centuries back: Dimensions from Digital Science and Microsoft Academic. In this study, we used publication data from these new databases and added publication data from two established databases (Web of Science from Clarivate Analytics and Scopus from Elsevier) to investigate scientific growth processes from the beginning of the modern science system until today. We estimated regression models that included simultaneously the publication counts from the four databases. The results of the unrestricted growth of science calculations show that the overall growth rate amounts to 4.10% with a doubling time of 17.3 years. As the comparison of various segmented regression models in the current study revealed, the model with five segments fits the publication data best. We demonstrated that these segments with different growth rates can be interpreted very well, since they are related to either phases of economic (e.g., industrialization) and / or political developments (e.g., Second World War). In this study, we additionally analyzed scientific growth in two broad fields (Physical and Technical Sciences as well as Life Sciences) and the relationship of scientific and economic growth in UK. The comparison between the two fields revealed only slight differences. The comparison of the British economic and scientific growth rates showed that the economic growth rate is slightly lower than the scientific growth rate.

# 1 Introduction

Growth of science is an ongoing topic in empirical and theoretical studies on science of science. In a recent overview of science of science studies, Fortunato et al. (2018) stated that "early studies discovered an exponential growth in the volume of scientific literature … a trend that continues with an average doubling period of 15 years". The investigation of growth processes leads to results that can be used to characterize science. For example, if the literature doubles every 15 years, science would be characterized by immediacy: "the bulk of knowledge remains always at the cutting edge" (Wang & Barabási, 2021, p. 163). Results on growth processes can also be used to investigate the validity of theories on the development of science: Does science follow a slow, piecemeal process or a process with normal science interrupted by revolutionary periods with an increased level of activity (Kuhn, 1962; Tabah, 1999)? Popular early studies on growth of science have been published by the theoretician of science, Derek John de Solla Price (1965; 1951, 1961) who can be seen as the pioneer in investigating growth of science processes (see de Bellis, 2009). According to Price (1986), the development of science follows the law of exponential growth: "at any time the rate of growth is proportional to the … total magnitude already achieved – the bigger a thing is, the faster it grows" (p. 4). Although empirical and theoretical studies in previous decades have confirmed exponential growth, a precise estimation of the growth rate based on reliable and sound publication data has not been done yet.

In most of the studies on growth of science published hitherto, bibliometric data have been used to measure growth of science (an alternative measure is the number of researchers, for instance). It is an advantage of using bibliometric data (compared to other data) that large-scale, multi-disciplinary databases are available based on worldwide publication productions. Another advantage is the characteristic of most scientific disciplines that publications are the main outcome: "science would not exist, if scientific results are not communicated. Communication is the driving force of science. That is why scientists have to publish their research results in the open, international scientific literature. Thus, publications are essential" (van Raan, 1999, p. 417). According to Merton (1988), "what we mean by the expression 'scientific contribution': an offering that is accepted, however provisionally, into the common fund of knowledge" (p. 620).

In a previous study (Bornmann & Mutz, 2015), two authors of the current study investigated the growth of science based on data from the Web of Science database (Clarivate Analytics; Birkle, Pendlebury, Schnell, & Adams, 2020). Bornmann and Mutz (2015) not only used annual publication numbers but also cited references data (see Marx & Bornmann, 2016, for an overview of the use of cited references data in scientometrics). They argued that Web of Science data (publication counts) are scarcely suitable to investigate early periods of modern science, since early publications are not sufficiently covered. Cited references may have the advantage of covering these early periods and a wider range of document types, including journal articles, books, book contributions or proceedings, which are still not fully included in the databases. However, cited references data can only serve as a less-than-ideal proxy of publication numbers, because non-cited publications are not considered. In recent years, new bibliographic databases have been introduced: Dimensions (Herzog, Hook, & Konkiel, 2020; Hook, Porter, & Herzog, 2018) from Digital Science and Microsoft Academic (Wang et al., 2020) which can be used to study growth processes in science from centuries back. Thus, it is the intention of the current study to use both databases for investigating these processes and compare the results with those from Web of Science and Scopus (Elsevier; Baas, Schotten, Plume, Côté, & Karimi, 2020).

With Dimensions, Microsoft Academic, Web of Science, and Scopus, we considered in this study (the most) important multi-disciplinary literature databases currently available. The comparison of the empirical results based on the four databases may point to an assessment of growth processes in science that might be interpreted as valid – since the assessments can be made independently of the use of single data sources. We investigated the growth processes not only for all annual publications in the databases, but also for two broad fields: (1) Physical and Technical Sciences and (2) Life Sciences (including Health Sciences). We selected these broad fields and did not consider further fields such as social sciences and humanities. Only for these two fields, we can be sure that publication data can be used as valid proxy for research activity.

In this study, we additionally undertook a comparative analysis of economic and scientific growth processes. According to Price (1986), the theoretical basis for the study of econometrics is similar to that for the study of scientometrics: both follows the law of exponential growth (differences lie in the parameters). Previous scientometrics research revealed that growth of science is related to economic development (Fernald & Jones, 2014; Salter & Martin, 2001). Although a national science system producing high-quality research is – without doubt – an important condition for national wealth, we primarily consider money as necessary input to the science system (and thus, economic growth as independent variable). In principle, national wealth can be achieved without a modern science system (as has been done for centuries), but (modern) science needs economy to exist and function.

Our comparative analysis of scientometrics and econometrics could not be done based on worldwide data, since long-time series for publication counts and economic growth indices are not available at this level. Following seminal research by May (1997) and King (2004a, 2004b) on the relationship of science and economy, we focus instead on UK for which time series of economic development are available that reach back to the 17th century (Thomas, Hills, & Dimsdale, 2010). Such historical data are not available for other countries (to the best of our knowledge). Using similar statistical methods as for publication data, we investigated in this study annual growth rates in gross domestic product (GDP) as a measure of economic wealth of a nation similar to the approach by King (2004a, 2004b).

## 2  Methods

### 2.1  Dataset used

We used bibliometric and economic data in this study. The five different databases and datasets are as follows:

**Web of Science**: The core citation indices of Web of Science (SCI-E, SSCI, and A&HCI) date back into the 1960s when they were founded by Eugene Garfield. The other citation indices were started later on (e.g., CPCI-S and CPCI-SSH). In total, the publications indexed in the Web of Science are divided into 44 different document types (e.g., review, news item or note). The coverage of the scientific literature dates back to 1900. The Web of Science is more selective with respect to the choice of indexed sources than the other databases in this study (Visser, van Eck, & Waltman, 2020). We used the advanced search of the Web of Science online interface[i] with the query "py=1900-2018" in the indices SCI-E, SSCI, A&HCI, CPCI-S, CPCI-SSH, BKCI-S, BKCI-SSH, ESCI, CCR-EXPANDED, and IC (Index Chemicus) (date of search: 30 August 2019). No restriction on document types was imposed. Via the "analyze results" function applied to publication years, we were able to download the number of indexed papers per year.

Broad subject categories were defined via the Web of Science subject categories:
- **Physical and Technical Sciences**: "Astronomy & Astrophysics", "Chemistry", "Crystallography", "Electrochemistry", "Geochemistry & Geophysics", "Geology", "Mathematics", "Meteorology & Atmospheric Sciences", "Mineralogy", "Mining & Mineral Processing", "Oceanography", "Optics", "Physical Geography", "Physics", "Polymer Science", "Thermodynamics", "Water Resources", "Acoustics", "Automation & Control Systems", "Computer Science", "Construction & Building Technology", "Energy & Fuels", "Engineering", "Imaging Science & Photographic Technology", "Information Science & Library Science", "Instruments &

Instrumentation", "Materials Science", "Mechanics", "Metallurgy & Metallurgical Engineering", "Microscopy", "Nuclear Science & Technology", "Operations Research & Management Science", "Remote Sensing", "Robotics", "Science & Technology Other Topics", "Spectroscopy", "Telecommunications", and "Transportation".

- **Life Sciences** (including Health Sciences): "Agriculture", "Allergy", "Anatomy & Morphology", "Anesthesiology", "Anthropology", "Audiology & Speech-Language Pathology", "Behavioral Sciences", "Biochemistry & Molecular Biology", "Biodiversity & Conservation", "Biophysics", "Biotechnology & Applied Microbiology", "Cardiovascular System & Cardiology", "Cell Biology", "Critical Care Medicine", "Dentistry, Oral Surgery & Medicine", "Dermatology", "Developmental Biology", "Emergency Medicine", "Endocrinology & Metabolism", "Entomology", "Environmental Sciences & Ecology", "Evolutionary Biology", "Fisheries", "Food Science & Technology", "Forestry", "Gastroenterology & Hepatology", "General & Internal Medicine", "Genetics & Heredity", "Geriatrics & Gerontology", "Health Care Sciences & Services", "Hematology", "Immunology", "Infectious Diseases", "Integrative & Complementary Medicine", "Legal Medicine", "Life Sciences Biomedicine Other Topics", "Marine & Freshwater Biology", "Mathematical & Computational Biology", "Medical Ethics", "Medical Informatics", "Medical Laboratory Technology", "Microbiology", "Mycology", "Neurosciences & Neurology", "Nursing", "Nutrition & Dietetics", "Obstetrics & Gynecology", "Oncology", "Ophthalmology", "Orthopedics", "Otorhinolaryngology", "Paleontology", "Parasitology", "Pathology", "Pediatrics", "Pharmacology & Pharmacy", "Physiology", "Plant Sciences", "Psychiatry", "Public, Environmental & Occupational Health", "Radiology, Nuclear Medicine & Medical Imaging", "Rehabilitation", "Reproductive Biology", "Research & Experimental Medicine", "Respiratory System", "Rheumatology", "Sport Sciences", "Substance Abuse",

"Surgery", "Toxicology", "Transplantation", "Tropical Medicine", "Urology & Nephrology", "Veterinary Sciences", "Virology", and "Zoology".

**Scopus**: Scopus was launched in 2004 by the publisher Elsevier. Coverage of the scientific literature dates back to 1861. The publications indexed in Scopus are divided into 16 different document types. Scopus has a broader coverage than Web of Science, especially in the social sciences and humanities (Visser et al., 2020). We used the advanced search of the Scopus online interface[ii] with the query "PUBYEAR AFT 1800" for this study (date of search: 30 August 2019). No restriction on document types was imposed. Via the "analyze search results" function applied to publication years, we were able to conveniently download the number of indexed papers per year.

Broad subject categories were defined via the Scopus subject areas:

- **Physical and Technical Sciences**: "Chemical Engineering", "Chemistry", "Computer Science", "Earth and Planetary Sciences", "Energy", "Engineering", "Environmental Science", "Materials Science", "Mathematics", and "Physics and Astronomy".

- **Life Sciences** (including Health Sciences): "Medicine", "Nursing", "Veterinary", "Dentistry", "Health Professions", "Multidisciplinary[iii]", "Agricultural and Biological Sciences", "Biochemistry, Genetics and Molecular Biology", "Immunology and Microbiology", "Neuroscience", and "Pharmacology, Toxicology and Pharmaceutics".

**Microsoft Academic**: Microsoft Academic was first released in 2016. It can be considered an unconventional bibliographic database because its content is not delivered by the publishers but found by the search engine Bing on the publisher's websites. This implies that especially the data from Microsoft Academic might have a bias towards publications with a digital footprint. However, many publishers provide websites for their older publications, too. Microsoft Academic offers a basic search interface[iv] and bulk data access via the Azure platform[v]. Microsoft Academic has a broader coverage than Web of Science and Scopus (Visser et al., 2020). We downloaded a snapshot of the Microsoft Academic data from the Azure platform (last update: 11 January 2019). The raw Microsoft Academic data were imported and processed in a locally maintained PostgreSQL database at the Max Planck Institute for Solid State Research. Our current snapshot of the Microsoft Academic database contains bibliographic data of 212,209,775 publications, such as title, publication year, and document type. Content coverage dates back to 1800. The publications indexed in Microsoft Academic are divided into five different document types ("Journal", "Patent", "Conference", "BookChapter", and "Book"). Unfortunately, 77,227,143 indexed items are not assigned to any document type. Via SQL commands, we produced items per publication year statistics for all items with known document type in the Microsoft Academic database excluding the document type patent but included the items without document type for a separate analysis.

Microsoft Academic offers a subject classification on different hierarchical levels. There are 19 different fields on the highest level. Broad subject categories were defined via that highest level:

- **Physical and Technical Sciences**: "Geology", "Chemistry", "Materials science", "Mathematics", "Engineering", "Environmental science", "Physics", "Geography", and "Computer science".
- **Life Sciences** (including Health Sciences): "Biology" and "Medicine".

**Dimensions**: Dimensions is the most recent database used in this study. It was launched in 2018 by Digital Science and contains meta-information about grants, publications, clinical trials, and patents. Like Web of Science and Scopus, Dimensions receives publication data information from the publishers but pursues a different indexing strategy. Dimensions tries to cover as many publications and publication types as possible. Dimensions is accessible via an online search interface[vi], an API, and, additionally, Digital Science shares the raw data without cost for research purposes[vii]. The raw Dimensions data (last update: 26 September 2019) were downloaded, imported and processed in a locally maintained PostgreSQL database at the Max Planck Institute for Solid State Research. The raw data of the Dimensions database are provided as separate sub-databases: "Grants", "Publications", "Clinical trials", and "Patents". In the following, by using the term "Dimensions" in the text, we refer only to the Dimensions sub-database "Publications". The indexed publications therein are divided into six different publication types ("article", "chapter", "proceeding", "preprint", "monograph", and "book"). Dimensions offers the second largest coverage of the literature in this study (Visser et al., 2020). Dimensions offers a much larger coverage of books and book chapters than Web of Science or Scopus (Clarivate, 2020; Elsevier, 2020; Taylor, 2020). Via simple SQLs, we produced publications per publication year statistics without restrictions on publication types in the Dimensions database.

Dimensions offers many different classification schemes, some of them are focused on specific disciplines or topic like Sustainable Development Goals (SDGs). For the purposes of our study, we have made use of the Dimensions implementation of the Australian and New Zealand Standard Research Classification (ANZSRC) Fields of Study (FOR) codes, as per the 2008 field definitions.[viii] The ANZSRC codes are delivered at three levels, the two least granular levels of which have been implemented in Dimensions. There are 22 fields of the highest level. Broad subject categories were defined in this study via that higher level:

- **Physical and Technical Sciences**: "Mathematical Sciences", "Physical Sciences", "Chemical Sciences", "Earth Sciences", "Environmental Sciences", "Information and Computing Sciences", "Engineering", "Technology", and "Built Environment and Design".
- **Life Sciences** (including Health Sciences): "Biological Sciences", "Agricultural and Veterinary Sciences", and "Medical and Health Sciences".

   **FRED**: The economic research department of the Federal Reserve Bank of St. Louis (FRED) offers a series of datasets for economic analyses and for analyses of the historical development of economic indicators. A time series from 1770 to 2016 of the annual "Nominal

Gross Domestic Product at Market Prices in the UK, Millions of British Pounds, Annual, Not Seasonally Adjusted" (NGDPMPUKA) for UK was downloaded as an Excel table.[ix] We use in the following the term "growth domestic product" or GDP instead of NGDPMPUKA (to facilitate the reading of the results). Since the values are nominal values, GDP is not adjusted for inflation. Publication counts for UK were retrieved from Dimensions for the years 1788 until 2016.

The data retrieved from the various databases is the number of publications published in one year. For the growth analysis, however, the cumulative number of publications is used. If, for example, up to a year x, 1,000 publications were published, and in the year x 100, the accumulated number of publications in the year x is 1100 publications. The difference to year x-1 is exactly the absolute growth in year x, i.e. 100 publications, the number of publications published in year x. For simplification, "number of publications" is used below instead of "cumulative number of publications".

## 2.2    Statistical analyses

Scientific growth processes do not necessarily run homogeneously over time, especially when a long-time horizon is chosen, for example, from the beginning of modern science in the 16th/17th century until today. Therefore, modern growth analysis has to simultaneously address three different problems: (1) Science can grow according to different growth functions which provide hypotheses about the nature of growth processes (e.g., unrestricted exponential). (2) It can be assumed that science grows at different rates in different time periods or segments, i.e., growth rates vary over time. (3) Growth functions might vary across different databases such as Scopus or Web of Science covering different time horizons. In the following sections, solutions to the three problems are presented which refer to growth functions (unrestricted and restricted exponential growth), segmented regression, and latent growth curve models.

**2.2.1 Growth functions**

The simplest growth function is that of unrestricted growth in the form of an exponential function, where the growth of science in each year is proportional to the volume of publications available in the previous year. An equal percentage of volume grows every year. For example, if we assume an annual growth rate of 10% and 100 publications in a certain year, then there are 100+0.10*100=110 publications in the following year. One year later, there are 110+0.10*110=121 publications (and so on). Another growth function assumed by Price (1963) is that of restricted growth: Science would run exponentially at the beginning, but with time the growth process approaches an upper capacity limit with constantly decreasing growth rates (s-shaped course). In view of the limited capacities of human and investment capital for research (and other sections of society), the latter thesis by Price (1963) seems to be more plausible than the simplest growth function: Since resources (human resources, capital) are limited, growth cannot be limitless either.

These considerations make it necessary to choose a statistical analysis approach that starts from different time segments, in which different growth rates apply and different growth functions are possible as well. The time segments themselves are not known in advance and have to be estimated. Such an opportunity is offered by the "segmented regression" or "piecewise regression" analyses, which start from different intervals of a dependent variable (in this case: time). These regression analyses apply different functional relationships and simultaneously make it possible to estimate time segments and parameters of the growth functions (Gallant & Fuller, 1973; McZgee & Carleton, 1970; Schwarz, 2015; Toms & Lesperance, 2003; Valsamis, Ricketts, Husband, & Rogers, 2019; Wagner, Soumerai, Zhang, & Ross-Degnan, 2002). In this study, we assume a time series in which the total number of publications $y_t$ is available per year, where $t$ denotes the index of the time series, and $t$=0 the starting year of the time series (e.g., for the year 1665: $t$=year-1665). We assume two growth functions (see above):

**Unrestricted exponential growth**

The functional relationship for exponential growth assumes that the derivative of the function is proportional to the function itself: $f(t) \sim b_1 f(t)$. The resolution of this differential equation leads to a functional relationship, which can be represented in the following statistical model:

$$y_t = f(t) = e^{b_0} e^{b_1 t} e^{\varepsilon_t} \qquad \varepsilon_t \sim N(\mathbf{0}, \sigma^2 \mathbf{CORR}_{\varepsilon_t \varepsilon_{t-1}}), \qquad (1)$$

where $e^{b_0}$ represents the initial volume of publications at the starting point of the time series ($t=0$), $b_1$ the growth constant, and $\varepsilon_t$ the residual with the variance $\sigma^2$ as well as the correlation matrix of the residuals $CORR_{\varepsilon t \varepsilon t-1}$. The latter is equated here with the identity matrix **I**, which means that the residuals do not (auto-)correlate. After the model estimation, we checked whether the residuals of the estimated model are actually auto-correlated or not. In the simplest case of an autoregressive process of first order (AR(1)), the residuals at time $t$ are (auto-)correlated with the residuals at time $t$-1.

If equation 1 is logarithmically transformed, a simple linear regression function can be obtained:

$$\ln_e(y_t) = b_0 + b_1 t + \varepsilon_t \qquad \varepsilon_t \sim N(\mathbf{0}, \sigma^2 \mathbf{CORR}_{\varepsilon_t \varepsilon_{t-1}}) \qquad (2)$$

The doubling time $k$ as the time the growth process needs to double the population size at a given time point is:

$$k = \ln(2) / (\ln(1+g)), \qquad (3)$$

where *k* is the doubling time and *g* is the growth rate. The *annual growth rate g* as the percentage change between two time points is $e^{b_1}-1$ for Eq. 1. For $b_1 = 0.05$, for example, *g* amounts to 0.051 or 5.1%.

**Restricted exponential growth (Verhulst-Pearl)**

For restricted exponential growth as a special case of a logistic growth model with a capacity limit *C*, the derivation of the function is proportional to the following function: $f(t)= b_1 f(t) (1-f(t)/C)$. The resolution of this differential equation leads to a functional relationship, which can be represented in the following statistical model (Tsoularis & Wallace, 2002, p. 28f.):

$$y_t = f(t) = \frac{e^K e^{b_0}}{(e^K - e^{b_0})e^{-b_1 t} + e^{b_0}} e^{\varepsilon_t} \qquad \varepsilon_t \sim N(0, \sigma^2 \mathbf{CORR}_{\varepsilon_t \varepsilon_{t-1}}) \quad (4)$$

It can be seen from equation 4 that if $t \to \infty$, the exponential expression in the denominator, $e^{-b_1 t}$, goes towards zero and the function approaches the capacity limit $C = e^K$. At time $t=0$, the start of the time series, the exponential expression in the denominator, $e^{-b_1 t}$, is equal to 1 and the function corresponds to the initial volume $e^{b_0}$ multiplied by the error term $e^{\varepsilon_t}$. A limited growth is assumed only for the first segment to rule out or not the de Solla Price hypothesis of growth of science. The combination of s-shaped segments over time seems to be implausible in light of the empirical results on the growth of science by Bornmann and Mutz (2015).

If equation 4 is logarithmically transformed, the following linear regression function results:

$$\ln_e(y_t) = K + b_0 - \ln_e((e^K - e^{b_0})e^{-b_1 t} + e^{b_0}) + \varepsilon_t \qquad \varepsilon_t \sim N(\mathbf{0}, \sigma^2 \mathbf{CORR}_{\varepsilon_t \varepsilon_{t-1}})$$
(5)

In the following, we call the "restricted exponential growth model (Verhulst-Pearl)" the "logistic growth model".

### 2.2.2 Segmented regression

Following classic theories of economic development, we consider the process of development in science and economy as a sequence of historical stages (Dang & Sui Pheng, 2015). In addition to the functional model, therefore, a statistical framework model is required. We used segmented regression which defines the regression models for different time segments and can be represented in the form of nested IF-THEN clauses for each segment $j$. In the case of unrestricted growth in all segments $j$, the following overall model applies with *year* $t_0$ as the starting year of the time series (e.g., 1665):

IF t ≤ α₁ THEN
$$\log(y_t) = b_0 + b_1(t - t_0) + \varepsilon_t$$
ELSE IF t ≤ α₂ THEN
$$\log(y_t) = b_0 + b_1(a_1 - t_0) + b_2(t - a_1) + \varepsilon_t$$
ELSE IF t ≤ a₃ THEN
$$\log(y_t) = b_0 + b_1(a_1 - t_0) + b_2(a_2 - a_1) + b_3(t - a_2) + \varepsilon_t$$
ELSE IF t ≤ α_j THEN

$$\log(y_t) = b_0 + (j>1)(\sum_{k=1}^{j-1} b_k(a_k - a_{k-1})) + b_j(t - a_{j-1}) + \varepsilon_t \qquad \varepsilon_t \sim N(\mathbf{0}, \sigma^2 \mathbf{CORR}_{\varepsilon_t \varepsilon_{t-1}}), \quad (6)$$

where $a_j$ denotes the year at which the $j^{\text{th}}$ time segment $j$ ends, and where $a_0 = t_0$ – the starting year of the time series. In addition to the parameters of the growth model, the year parameters $a_1$ to $a_{j-1}$ are estimated. The same distribution of residuals is assumed for each segment.

Publication counts is a count variable. The variable includes positive integer values with zero. This implies that the values are distributed, for example, according to a Poisson distribution (Hilbe, 2014, p. 2). In this study, however, a logarithmic transformation (base *e*) of the publication data was favored over a Poisson model for the following reasons: (1) with regard to growth rates of science, unrestricted growth can be assumed, in which the logarithmic transformation leads to a simple linear regression function. The parameters of the function can be interpreted in terms of the original non-transformed growth function (Panik, 2014, p. 33). (2) If it can be demonstrated that the observed values are well explained by the function (because of low residual variance), then neither the distribution function nor the transformation play a major role. (3) Due to the smaller scale of the values resulting from log-transformation, there is a greater chance that complex statistical models converge in the estimation process.

### 2.2.3 Piecewise latent growth curve model with missing imputation

In this study, we used data from several bibliographic databases. We therefore needed to find an answer to the question of how the various datasets reflecting the same information (scientific output) should be analyzed statistically. It was one option to conduct the analyses for each database separately. This approach would accord with the analyses by Bornmann and Mutz (2015). Analyses for each database separately, however, run the risk of obtaining four different results that might reflect specific aspects of a database. Another option was to analyze the data from the different databases within one statistical model. This solution would still need solutions to the following problems:

(1) The time intervals at which publication data are available vary from database to database. The largest time interval (from 1665 to 2018) is available from Dimensions. To analyze only the time interval for which all databases provide complete data would significantly limit the period of investigation of the development of science. (2) The publication data vary greatly in volume between the databases. Dimensions, for example, has the highest volume of publications when the entire time series is considered, whereas Web of Science has the comparatively lowest volume. Here, the question arises whether some form of data weighting according to volume is necessary.

The solution for these problems that we favored in this study was the application of the so-called "Latent Piecewise Growth Curve Model". This model can be run in conjunction with an approach based on completed time series, i.e. incomplete time series are treated statistically as missing value problem. Another possible solution for the problems would be to refer only to those years for which all time series have information. This solution would limit the time horizon of the analysis (elimination of epochs). Furthermore, the possibility of looking further into the past would get lost with consideration of only complete information. The problem of missing values only becomes relevant when the years before the turn into the 20th century are considered.

There are several methods available to deal with missing values (Little & Rubin, 2019). The most important are two types of procedures: "Maximum Likelihood" and "Imputation". Maximum Likelihood methods can be used to identify different patterns of missing values and then efficiently estimate the parameters in the estimation procedure using all available information across the patterns, so called "Full Information Maximum Likelihood" (FIML). In imputation procedures, missing values are replaced by estimated values, for example by the mean value of a variable. In the "multiple imputation procedure" several predicted values from a stochastic regression on variables with full information (here time series) are used for a missing value, representing the uncertainty in the estimate.

Three different assumptions about the missing-value process are crucial for both procedures. In the case of "Missing completely at random" (MCAR), it is assumed that the missing value process is completely random, i.e., the missing values do not dependent on observed values of other variables or the unobserved values of the variable under investigation itself. The missing value process can be ignored. A case-wise deletion would be appropriate in this case. "Missing at random" (MAR) assumes that the missing value process depends on observed values of other variables, but not on the unobserved values of the variable under investigation itself. "Missing not at random" (MNAR) assumes that the missing value process depends not only on observed values of other variables in the data but also on unobserved values of the variable under investigation itself.

Imagine, for example, database providers would exclude papers of specific publication years, because only a small set of documents were published. An MNAR assumption is not very plausible in this case. Database providers did not make any selections of publication years on the basis of the number of publications in a year. Since bibliographic databases provide measurement replications of the same growth process with high correlations (>.90) among the time series, the MCAR assumption cannot hold either, leaving the assumption of "Missing at Random". Our data from Web of Science, for example, cover the range from 1900 to 2018, and our data from Dimensions cover the range from 1670 to 2016. MAR requires that the missing values of Web of Science between 1670 and 1899 are not the results of intended actions by the database provider, Clarivate Analytics. In the case of intended actions, for example, the company would systematically (completely) leave out publication years with low publications counts. Due to the fact that the missing value process is not observable, unfortunately, MAR cannot be verified.

We opt for a multiple imputation procedure. In contrast to FIML, the procedure allows to make imputed values visible in order to check for possible biases. The problem of missing values becomes relevant in this study when we focused in time before 1900. The assumed inaccuracy of the model estimation by missing imputation reflects the uncertainty of the historical perspective: the further the empirical analysis goes back in history, the more uncertain the results become.

In the *first step* of the imputation procedure in this study, based on the complete information across all four time series / databases, the missing values of a time series are imputed with estimated values. To take into account the inaccuracy of values in the estimation (when imputed values are used), five imputed values are estimated for each missing value, which should represent a random sample of missing values. Graham, Olchowski, and Gilreath (2007) recommend more than 5 imputed values. The relative efficiency of an imputation estimator as a measure of how well the true parameters in the population are estimated was very high (above .99). The statistical estimation of one imputation was too time-consuming to allow for more imputations. A Markov-Chain Monte Carlo (MCMC) procedure was used to estimate the imputed value from the available time series with full information.

In the *second step* of the imputation procedure in this study, for each of the five complete datasets with imputed values, a segmented regression model is estimated and then synthesized to an overall result considering the inaccuracy of the missing imputation in the calculation of standard errors. The point estimate of the overall segmented regression model parameter is the average of the parameters of the 5 complete-data estimates. The point estimate of the predicted value (missing or not) for each time point and database is the average of the corresponding estimates of the 5 complete-data estimates. The standard error of the overall segmented regression model parameter consists of two estimates: the average of the standard errors of the regression parameter estimated for each of the 5 imputed data sets (within variance) and the variability of the regression parameter across the 5 imputed data sets (between variance).

The main challenge in the analyses was to obtain convergence of the estimation algorithm across all models and all imputations. Especially for models with many segments, convergence problems occurred due to different scaling of the variance components (high variability in the intercept and decreasing variability in the slopes with increasing number of segments). Therefore, random effects were partly scaled (e.g., multiplied by 100 or 0.01) to establish convergence.

The statistical analyses in this study were done with the statistical software package SAS and the procedures PROC NLMIXED, PROC NLIN, PROC MI, and PROC MIANALYZE (SAS Institute Inc., 2015).

# 3 Results

In this section, the results of the model estimations are presented. The first five years of each time series were discarded for the estimations because they seemed to reflect only a pseudo segment or artifact without any empirical meaning. Therefore, the actual starting years were 1670 for Dimensions, 1805 for Microsoft Academic, 1905 for Web of Science, and 1866 for Scopus. Each time series ran until the year 2018.

### 3.1 Model comparison

Statistical model comparisons make it possible to rule out unrealistic models with poor model fit in order to get the model with the relatively best fit to the data. The model formulation is associated with certain assumptions about scientific growth (see Table 1): (1) A model with unconstrained *exponential growth* can be distinguished from a model with *logistic growth*. (2) One can distinguish whether the models based on different bibliographic databases come to similar or different results (are there *mixed-effects or not*?). (3) If there are significant differences between the results based on the databases, the following question would arise: Do the databases with a comparable high (low) volume of publications in the beginning of the time series show a high (low) increase in the later publication count? If so, the covariance or correlation between starting volume of publications and slope across the databases would be high (is there *covariance or not*?). (4) The models can provide different answers to the question of how many segments exist in the growth of science (*how many segments* can be distinguished?).

Model $M_1$ "Exponential growth" (see Table 1), for example, includes three parameters: intercept, slope and residual variance. If intercepts and slopes are allowed to vary across the four databases, two variance components were additionally estimated with overall five parameters. In $M_3$, the covariance of intercept and slope only for the first segment was added as a further parameter.

Instead of statistical significance testing, model comparison is undertaken in this study based on the Schwarz`s Bayesian information criterion (BIC). The smaller the BIC, the better the model fits the data (see Table 1). Models represent overall hypotheses about the nature of growth (e.g., exponential). The BIC is corrected for the number of parameters. A selection of models (e.g., number of segments) was made that were still estimable given the number of parameters and that still showed model improvement in terms of BIC.

Comparing model 1 and model 2, it becomes clear that a simple fixed-effects model ($M_1$) does not fit the data well. The differences between the growth curves based on the various databases are too large, so that a mixed-effects model ($M_2$) can be assumed which results in a significantly smaller BIC. The hypothesis of logistic growth can be rejected as well since the exponential model fits better. Among the models in Table 1, model $M_9$ with five segments and a covariance of intercept and slope in the first segment fits best for Physical and Technical Sciences, model $M_8$ with four segments fits bests for "all publications" and Life Sciences with a negligible improvement for "all publications" with 5 segments (period of 2nd World War). This result applies to all datasets (databases) considered in this study that refer to (1) all publications, (2) Physical and Technical Sciences publications, and (3) Life Science publications. Since the explained variance – measured in terms of the coefficient of determination ($R^2$) – exceeds .99, any autocorrelation among residuals or possible heterogeneity of residual variances can be neglected (equations 2 to 5). The covariance matrix of the residuals, **CORR**$_{\varepsilon_t,\varepsilon_{t-1}}$, is assumed to be an identity matrix **I**.

The model comparison in Table 1 demonstrates that the assumption of constant scientific growth over time is not realistic; hence, we can start with the premise that periods with different growth rates exist. This premise seems reasonable since, for example, the history of the 20$^{th}$ century is characterized by two World Wars with drastic consequences for the science system worldwide. As the results by Bornmann and Mutz (2015) based on cited references data have shown, the negative effects of the World Wars on scientific activities are clearly visible (for the estimated parameters of the model, see Table S1 in the Supplementary Information). Comparing model $M_3$ with model $M_2$ and model $M_6$ with model $M_5$, BIC improves in both cases. There is a covariance across all databases between the intercept and the slope in the first segment that is negative in all models. The higher the initial time series level of a database, the more the time series slope in the first segment is below the average slope of all databases et vice versa.

Table 1. Model comparison using Schwarz's Bayesian information criterion (BIC) for publication data from different bibliographic databases including all publications, Life Sciences publications, and Physical and Technical Sciences publications

| $M_{nr}$ | Model description | Mixed effects? | Covariance components? | Number of segments | Number of parameters | All publications | Life Sciences | Physical and Technical Sciences |
|---|---|---|---|---|---|---|---|---|
| $M_1$ | Exponential growth | No | No | 1 | 3 | 5155.60 | 3610.69 | 4864.45 |
| $M_2$ | Exponential growth | Yes | No | 1 | 5 | 2038.73 | 2711.35 | 3367.32 |
| $M_3$ | Exponential growth | Yes | Yes | 1 | 6 | 2018.60 | 2693.45 | 3346.32 |
| $M_4$ | Logistic growth | Yes | Yes | 1 | 6 | 2206.11 | 2723.64 | 3331.06 |
| $M_5$ | Segmented Regression to $M_2$ | Yes | No | 2 | 8 | 77.37 | 2324.49 | 3262.69 |
| $M_6$ | Segmented Regression to $M_3$ | Yes | Yes | 2 | 9 | 67.16 | 2308.77 | 3251.81 |
| $M_7$ | Segmented Regression to $M_3$ | Yes | Yes | 3 | 12 | -359.79 | -190.30 | 365.52 |
| $M_8$ | Segmented Regression to $M_3$ | Yes | Yes | 4 | 15 | -549.85 | -315.55 | 75.55 |
| $M_9$ | Segmented Regression to $M_3$ | Yes | Yes | 5 | 18 | -549.89 | 1478.02 | 12.27 |

With respect to the single time series of the GDP, a model with seven segments fit the data best (see Table 2). For publication counts, a model with eight segments shows the best fit (see Table 2). We additionally compared the models using the mean square error (MSE) and the BIC derived from the MSE to select certain models (Kim & Kim, 2016) (see Table S2 in the Supplementary Information for the estimated parameters of the model).

Table 2. Model comparison using Schwarz's Bayesian information criterion (BIC) for publication data and growth domestic product data (GDP) of UK

| $M_{nr}$ | Model description | Number of segments | Number of parameters | Publication count | | GDP | |
|---|---|---|---|---|---|---|---|
| | | | | MSE | BIC | MSE | BIC |
| $M_1$ | Exponential growth | 1 | 2 | 0.152 | -419.82 | 1.056 | 27.62 |
| $M_2$ | Logistic growth | 1 | 2 | 0.195 | -365.83 | 2.0 | 237.81 |
| $M_3$ | Segmented Regression to $M_1$ | 2 | 4 | 0.150 | -414.51 | 0.058 | -881.90 |
| $M_4$ | Segmented Regression to $M_1$ | 3 | 6 | 0.006 | -1,157.17 | 0.051 | -912.71 |
| $M_5$ | Segmented Regression to $M_1$ | 4 | 8 | 0.005 | -1,174.78 | 0.038 | -995.51* |
| $M_6$ | Segmented Regression to $M_1$ | 5 | 10 | 0.004 | -1,197.21* | 0.038 | -983.99* |
| $M_7$ | Segmented Regression to $M_1$ | 6 | 12 | 0.002 | -1,354.62* | 0.024 | -1,121.85 |
| $M_8$ | Segmented Regression to $M_1$ | 7 | 14 | 0.002 | -1,350.37 | 0.012 | -1,343.92 |
| $M_9$ | Segmented Regression to $M_1$ | 8 | 16 | 0.002 | -1,356.72 | 0.029 | -1,042.64* |
| $M_{10}$ | Segmented Regression to $M_1$ | 9 | 18 | 0.002 | -1,337.75* | 0.016 | -1,225.21 |

Notes: BIC = Schwarz's Bayesian information criterion, MSE = Mean square error, optimal models are grey shaded.
*No convergence of the iterations in the estimation process.

## 3.2 Growth rates of science (all publications)

In our analyses of growth processes in science using publication data, we follow typical assumptions such as those formulated by Long and Fox (1995): "while research productivity is not strictly equivalent to publication productivity, publication is generally taken as an indication of research" (p. 51).

Figure 1 shows the result of the unrestricted growth ($M_1$) and segmented unrestricted growth ($M_9$) models based on the data from Dimensions, Microsoft Academic, Scopus, and Web of Science. The graphs in the figure present the annual logarithmized number of publications cumulated across time. The grey dots represent the missing imputed values for one imputation, the colored symbols the observed values (the raw data from the databases), and the black solid line (with the two black dashed lines) the predicted values from the regression analyses (with 95% prediction intervals). As the results of the unrestricted growth ($M_1$) in Figure 1a show, the overall growth rate amounts to 4.10% with a doubling time of 17.3 years.

As the model comparison in section 3.1 revealed, a model with five segments fits the data best. The results of this model are presented in Figure 1b. The colored dashed lines show the individual regression line based on the data from the various databases, and the black solid line the overall regression for the whole data (across all databases). The symbols represent single values, either observed (colored symbols) or imputed (grey dots). The results in the figure show – with the exception of the results based on the Scopus data for the first segment – that the predicted values from the regression (dashed lines) cover the observed values (points) very well.

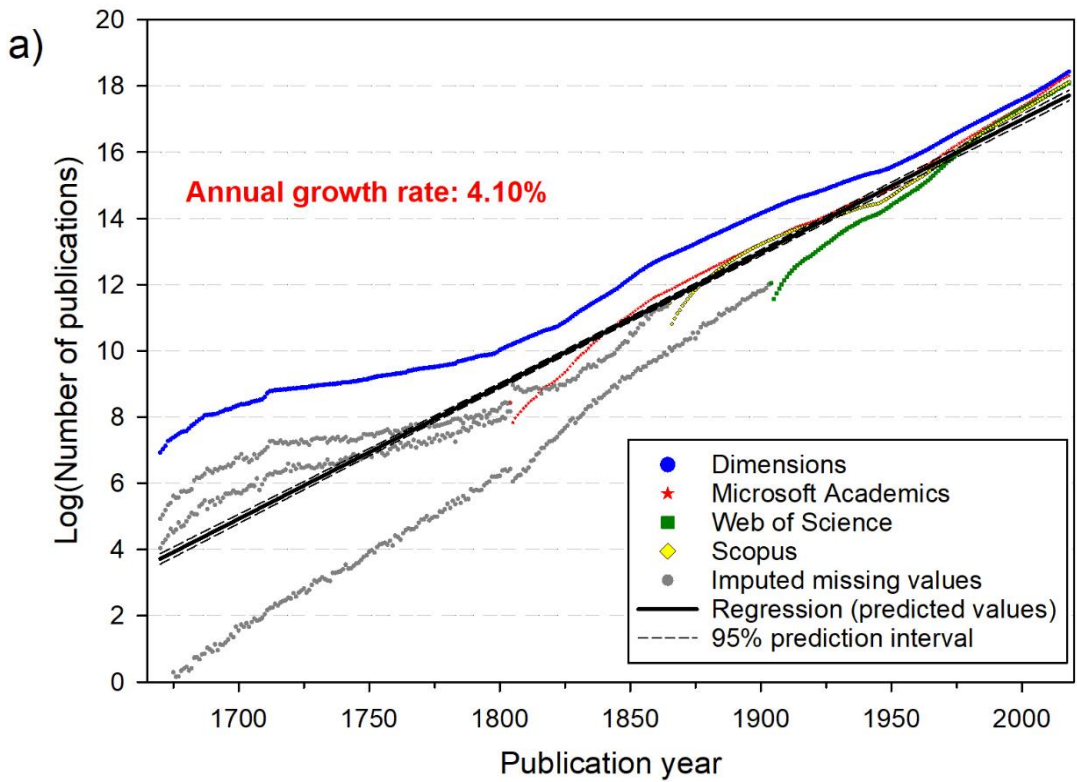
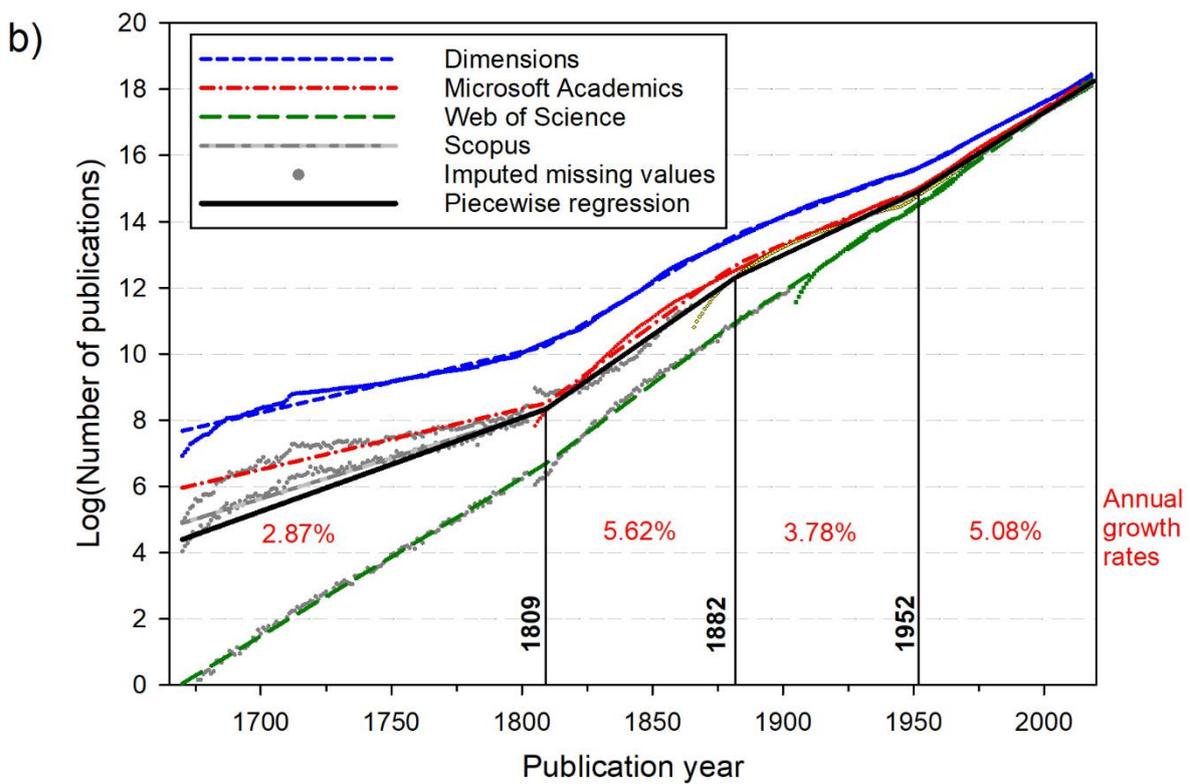

Figure 1. Plots for a) unrestricted growth ($M_1$) and b) segmented unrestricted growth ($M_9$) based on the number of publications from four bibliographic databases

The four segments in Figure 1b seem to represent separate historical epochs in the modern history of science: These segments with different growth rates are oriented towards either phases of economic (e.g., industrialization) and / or political developments (e.g., World Wars):

1. Phase: *Emergence of modern physics and pre-industrialization* (1675-1809). The phase up to the end of the Napoleonic wars is characterized by a moderate annual growth of 2.87% and a doubling time of 24.5 year, i.e., during 24.5 years the volume of publications doubles. This early phase of science is characterized by major discoveries in physics by Isaac Newton (1643-1727) and the development of the steam engine (James Watt, from 1769).

2. Phase: *Industrial Revolution* (1815-1881). In this phase of industrial revolution, science grew very strongly with an annual growth rate of 5.62% and a doubling time of 12.6 years.

3. Phase: *Economic crises and periods of World Wars* and Post-war (1881-1952): The development of science flattened out with an annual growth rate of 3.78% and a doubling time of 18.7 years. In this period, two economic depressions and two World Wars took place. The "long depression" is a period that started in 1873 and ended in 1896. The period is mainly characterized by a deflation in the USA and Europe (Capie & Wood, 1997). The "long depression" can be distinguished from the "Great Depression" that ranged from 1929 until the beginning of the Second World War.

4. Phase: *Post-war period* (after 1952 until today): Since 1952, science has grown exponentially without restrictions with an annual growth rate of 5.08% and a doubling time of 14.0 years.

In the statistical analyses of Microsoft Academic data, we considered all publications with known document types except patents, i.e., we excluded publications with unknown document type. Among publications without document type but with DOI, we identified book chapters and journal publications as well as conference papers and technical reports. We also found summaries and reports about conferences. Since not all publications can be seen as equal contributions to scientific progress, we analyzed the influence of the document type on our results by including documents with known and unknown document type without patents (see the results in the Supplementary Information, Figure S6). The differences between the results including all documents and only those documents with known document types are small. For all documents a further segment could be identified, which represent the period of Second World War (1940-1945).

### 3.3 Growth rates of science for Life Sciences and Physical and Technical Sciences

In addition to the analyses including all publications, we have also conducted analyses for two broad fields: Life Sciences and Physical and Technical Sciences. The estimated parameters of the models are reported in Table S1 in the Supplementary Information. The results are visualized in Figure 2 and Figure 3. With the comparison of two broad fields, we wanted to find out whether different fields are characterized by similar or different growth rates in their historical developments. As the results in Figure 2a show, the overall annual average growth rate for Life Science amounts to 5.07% with a doubling time of 14.0 years. The results for the Physical and Technical Sciences are similar, with a growth rate of 5.51% (see Figure 3a) and a doubling time of 12.9 years.

In agreement with the results for all publications in Figure 1b, the predicted values of the segmented regression model (dashed lines) cover the observed values (points) very well (high amount of explained variance) in both broad fields (see Figure 2b and Figure 3b). In both figures, we can observe trends that – although not completely congruent with the trends based on all publications – roughly illustrate the four central stages in the development of science and society: pre-industrialization (until 1793/1808), Industrial Revolution (till 1810

/1848), Second World War (1936-1943) only for Physical and Technical Science with a decline in the volume of publications, and the post-World War period.

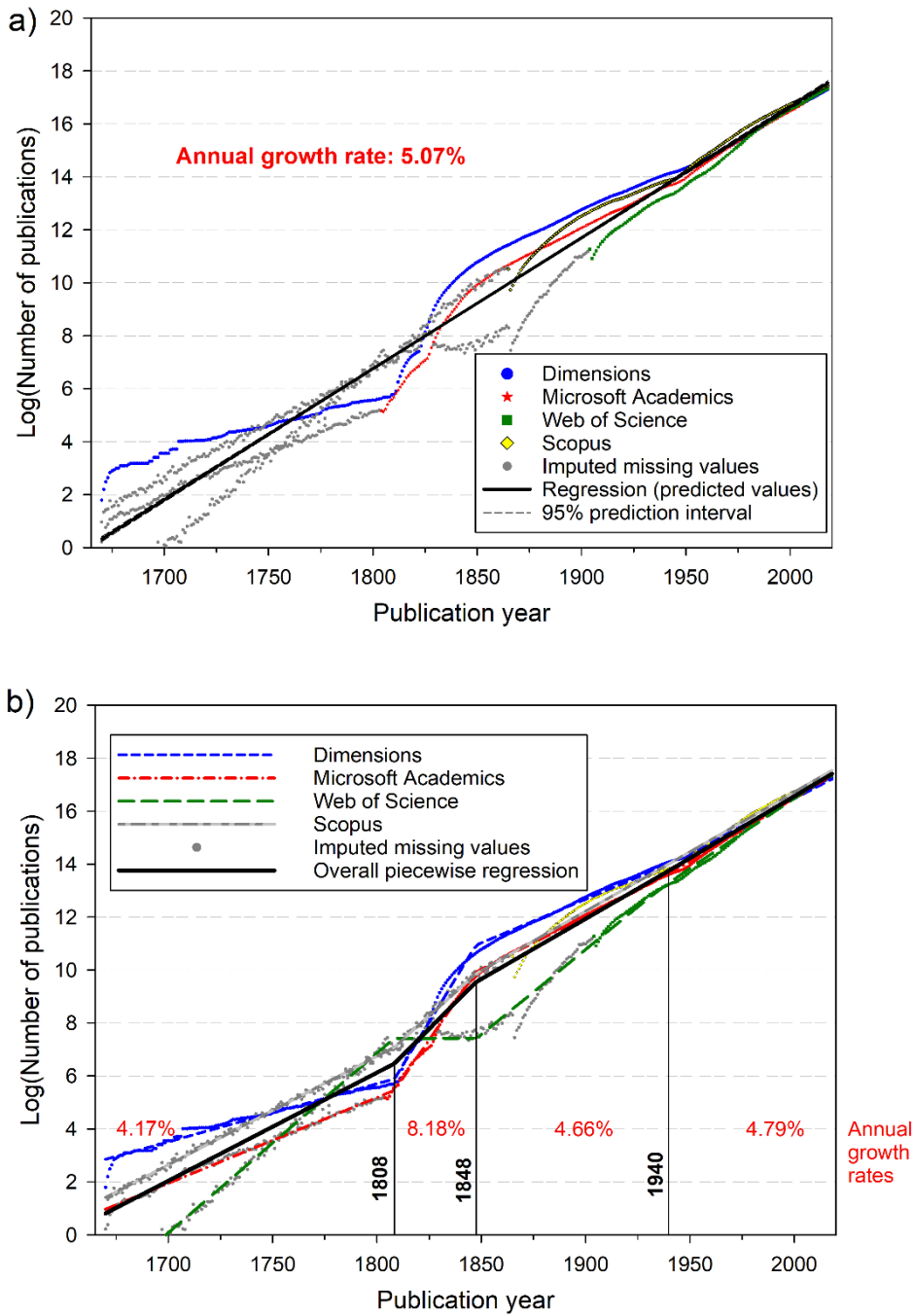

Figure 2. Plots for a) unrestricted growth ($M_1$) and b) segmented unrestricted growth ($M_9$) based on the number of publications in Life Sciences

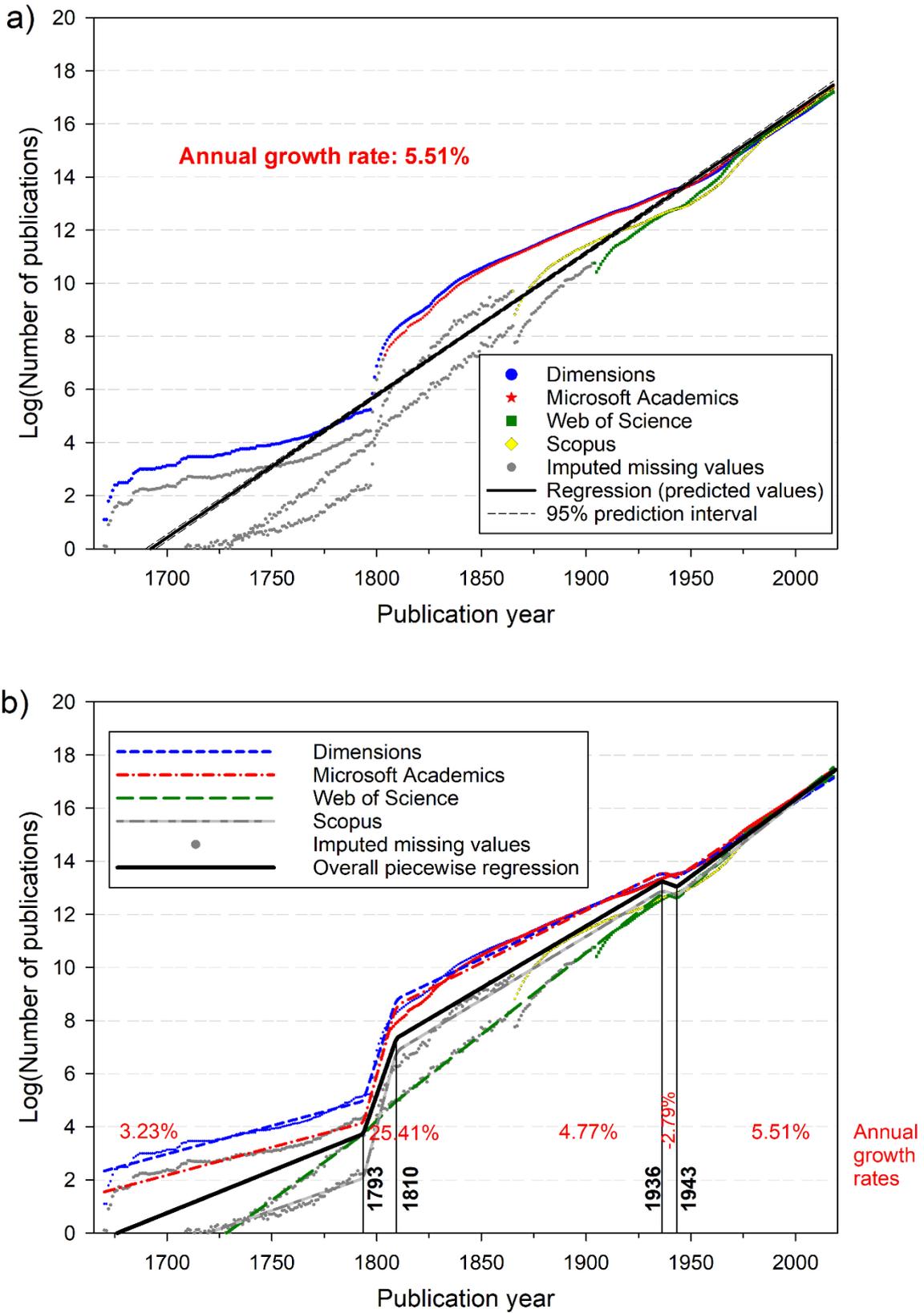

Figure 3. Plots for a) unrestricted growth ($M_1$) and b) segmented unrestricted growth ($M_9$) based on the number of publications in Physical and Technical Sciences

In the segment reflecting the period after 1945, with an annual growth rate of 5.51% and a doubling time of 12.9 years, the growth in the Physical and Technical Sciences is higher than the growth rate in the Life Sciences. In the Life Sciences the growth rate is 4.79% with a doubling time of 14.8 years. The growth rate in the Physical and Technical Sciences is also (slightly) higher than the growth rate that we calculated based on all publications in this segment (see Figure 1b): 5.08%.

### 3.4 Comparative analysis of growth rates of science and of growth domestic product in UK

For a comparative analysis of economic and scientific growth (using similar statistical methods), we used data from UK as explained in section 1. We analyzed logarithmic transformed GDP and logarithmic transformed cumulative publication data to estimate the different segments of growth rates and the growth rates themselves. Both rates are percentages and can be directly compared. The publication counts were obtained by the Dimensions database. The average annual growth rate of science in UK since 1780 is 4.97% (see Figure 4a). This corresponds to a doubling time of 14.64 years. This annual growth rate is slightly higher than the average worldwide growth rate of 4.10% (see Figure 1a). The statistical analysis revealed eight segments with different growth rates (see Figure 4b). The growth is, therefore, more differentiated than the overall growth with five segments (see Figure 1a). Between 1780 and 1805 (pre-industrialization) as well as 1805 and 1844 (early industrialization), a strong growth of 7.73% and 5.93%, respectively, can be observed. The growth weakens to 3.70% in the phase of industrialization from 1848 and the First World War as well as the 1920s.

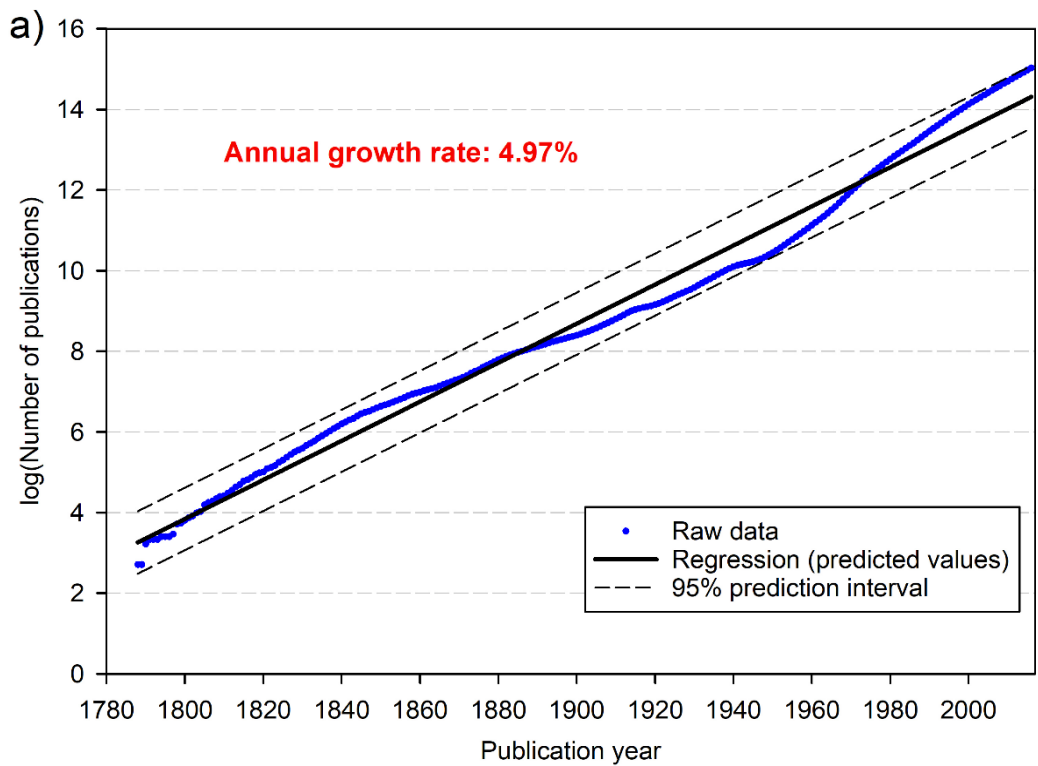

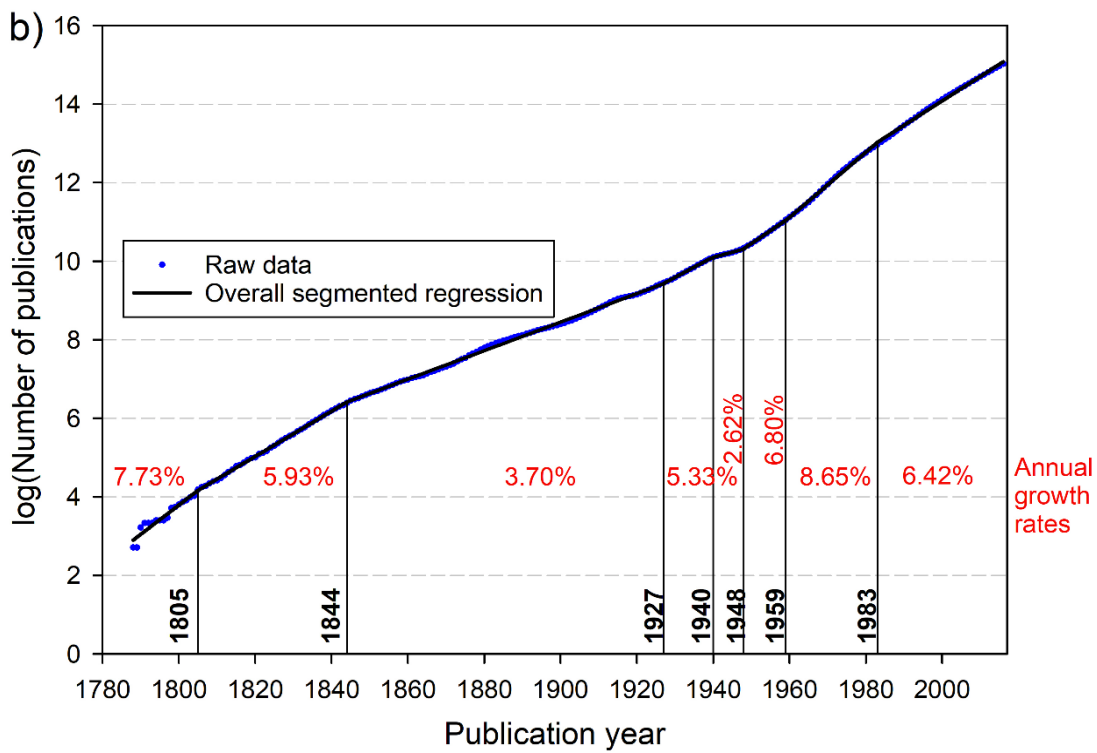

Figure 4. Plots for a) unrestricted growth ($M_1$) and b) segmented unrestricted growth ($M_9$) based on the number of publications from UK (using Dimensions data)

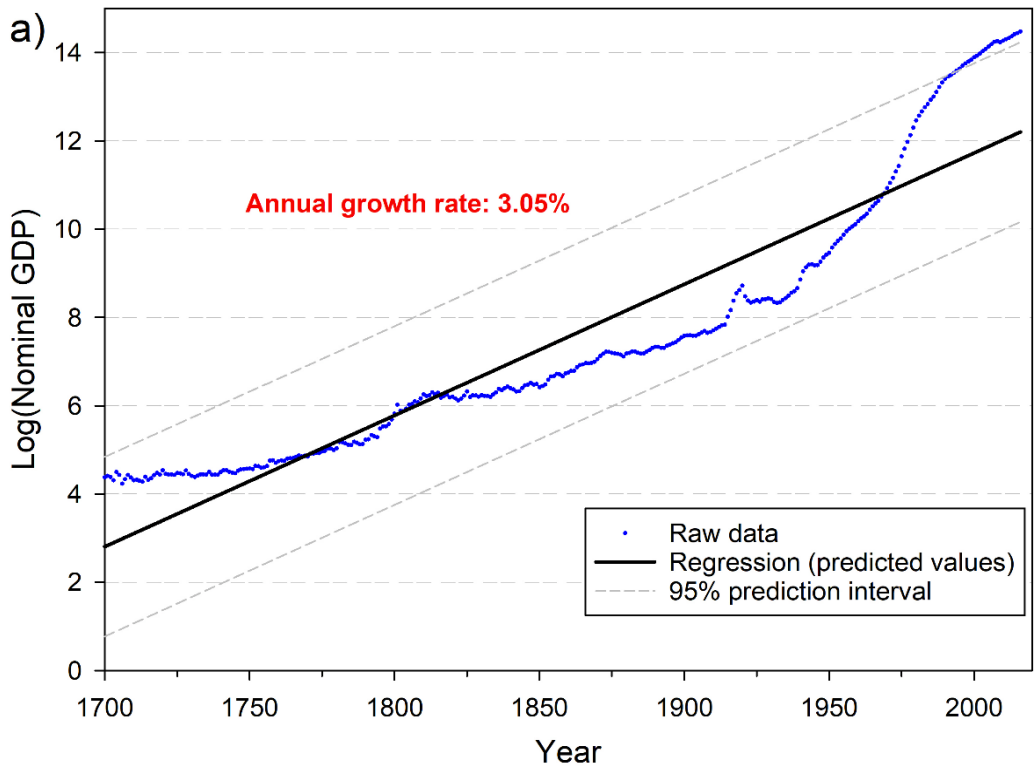

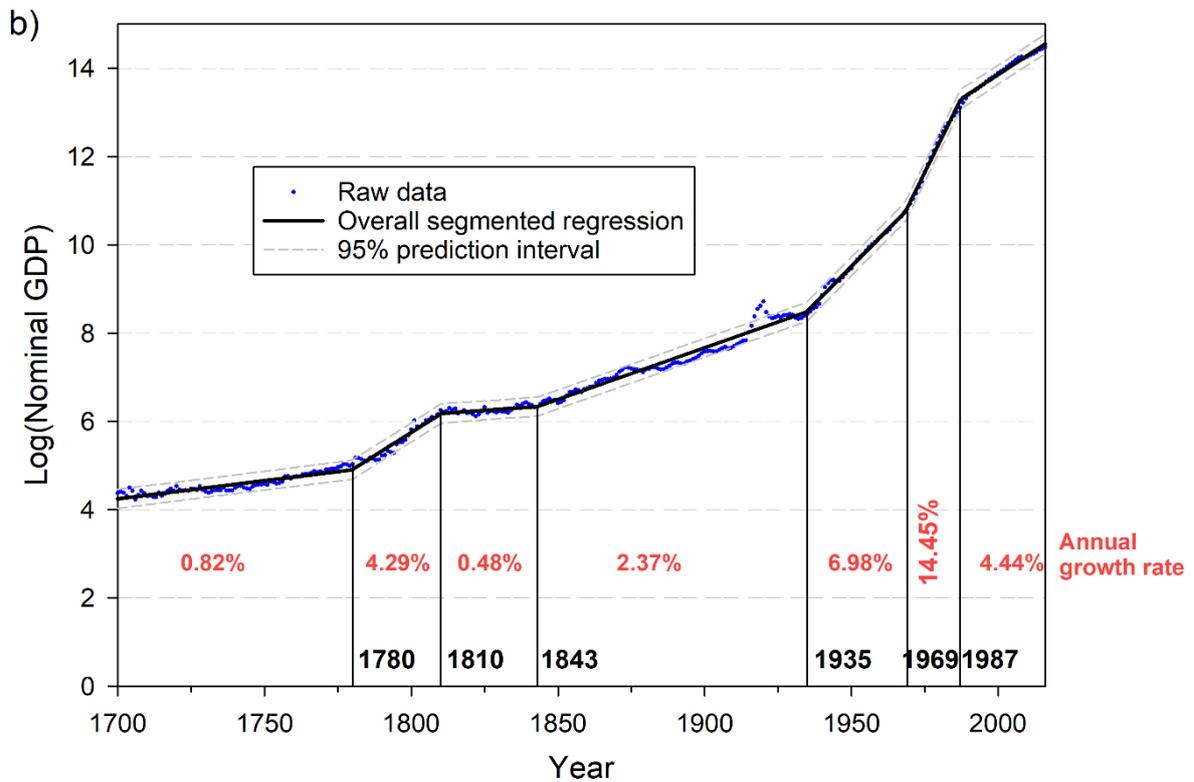

Figure 5. Plots for a) unrestricted growth ($M_1$) and b) segmented unrestricted growth ($M_8$) based on gross domestic product (GDP) data from UK (source: FRED Economic Research)

Comparable to worldwide results (see Figure 1b), a significant slowdown in scientific growth with a growth rate of 2.62% is apparent around the Second World War (between 1940 and 1948). While the overall analysis shows an unrestricted exponential growth after 1945 (see Figure 4b), the growth of science in UK took place in three stages: a strong growth of 6.80% until 1959, which intensified between 1959 and 1983 (8.65%), and slowed down to 6.42% in the years after 1983. The growth rates in these three segments are even higher than the worldwide growth rate of 5.28% in the corresponding time segment (between 1945 and 2018). At the beginning of the 1980s, Margaret Thatcher was Prime Minister of UK and with her party, the Conservative Party, having won the majority in the House of Commons for the second time in 1983.

Figure 5 shows the annual GDP growth rate between 1700 and 2018 for comparison with the publication numbers. The figure is based on the logarithmized annual GDP, presented as raw data and predicted values from the regression model. Previous studies investigating the relationship between economic and scientific growth have demonstrated positive relationships (e.g., Halpenny, Burke, McNeill, Snow, & Torreggiani, 2010; Hart & Sommerfeld, 1998; Ntuli, Inglesi-Lotz, Chang, & Pouris, 2015). The results in Figure 5a reveal an annual GDP growth rate of 3.05% and doubling time of 23.5 years which is lower than the growth rate based on publication counts of 4.97% (see Figure 4a).

At first glance, economic growth and scientific growth do not seem to be linked necessarily. A more detailed view shows, however, that both growths are related at certain points over time (see Figure 4b and Figure 5b). For example, science and economy grew from 1780 to the beginning of the 19th century (1810, 1805), i.e. in the phase of pre-industrialization at a comparable rate: whereas the economy grew by 4.29%, science grew by 5.93%. Furthermore, there is a coupling of economic and scientific development at the beginning of industrialization in the 1840s (1843, 1844) with a moderate annual growth rate of 2.37% in economy and 3.70% in science. A last temporal coupling can be observed in the years after the Second World War with a strong economic growth, especially from 1969 to 1987 of 14.45%. Three years later, in 1990, Margaret Thatcher resigned as Prime Minister. While the slowdown in the economy did not begin until after 1987, science began to grow at a rate of only 6.42% as early as 1983.

# 4 Discussion

Modern science is based on knowledge-producing institutions and processes (Gieryn, 1982). Current research is a method of "systematically exploring the unknown to acquire knowledge and understanding. Efficient research requires awareness of all prior research and technology that could impact the research topic of interest, and builds upon these past advances to create discovery and new advances" (Kostoff & Shlesinger, 2005, p. 199). Society expects a steady increase in scientific growth since only considerable growth processes would lead to growth in other sectors of society such as economics and health. Since (public) investments in science are frequently justified on the basis of growth of science and science contribution to national economic growth (Wagner, Park, & Leydesdorff, 2015), measurements of scientific growth processes are ongoing topics. These measurements are usually based on numbers of publications, since the results of research mostly appear in publications: "in academic institutions, publications constitute in all scientific-scholarly subject fields an important form of academic output" (Moed, 2017, p. 63). The results of Digital Science (2016) show that especially the journal article becomes increasingly popular as a medium for presenting scientific results. The popularity of journal articles could also be the consequence of the higher than average growth in disciplines using journal articles.

The motivation by researchers for publishing their results (in journal articles) is especially fostered by the specificity of the scientific reward system: "Publications have another function as well [besides the open availability of research results]: The principal way for a scholar to be rewarded for his contribution to the advancement of knowledge is through recognition by peers. In order to receive such an award, scholars publish their findings openly, so that these can be used and acknowledged by their colleagues" (Moed, 2017, p. 62). Although the publication of findings is so basic in science, researchers also process their findings in other forms of output (e.g., patents or presentations). An overview of indicators for measuring productivity based on these other forms can be found in Godin (2009). The problem of most of these indicators for measuring productivity or scientific growth, however, is that annual and historical data without missing values are scarcely available.

In this study, we used publication data from four literature databases to investigate scientific growth processes from the beginning of the modern science system until today. In accordance with the law of exponential growth, the results of the unrestricted growth show that the overall growth rate amounts to 4.10% with a doubling time of 17.3 years. This annual growth rate (over the various databases) is different from the Web of Science growth rate of 2.96% reported in Bornmann and Mutz (2015), since we considered in the current study a significantly longer time period than Bornmann and Mutz (2015): from 1900 until 2018 in this study (119 years) versus from 1980 until 2012 (33 years) in Bornmann and Mutz (2015). As the comparison of various segmented regression models in the current study revealed, the model with five segments fits the data best. We demonstrated that these segments with different growth rates can be interpreted very well since they are related to either phases of economic (e.g., industrialization) and/or political developments (e.g., World Wars). Obviously, the war efforts (allocation of funds) led to a visible decline in research (by output measure of publication) but research went on nevertheless, possibly with even more vigor. However, that research was not being made available openly for security reasons (and researchers pulled in for the sake of war efforts from physics to languages, material science to mathematics/emerging computer science) – and arguably the results of war-time research triggered post-war discoveries, too.

We additionally undertook two further analyses focusing on (1) growth in two broad fields (Life Sciences and Physical and Technical Sciences) as well as (2) the relationship between scientific and economic growth. (1) The comparison between the two broad fields revealed that although slight differences are observable, these differences are not so great that they can be denoted as fundamental. For example, whereas the overall annual average growth rate for Life Science is 5.07% with a doubling time of 14.0 years, the overall growth rate for Physical and Technical Sciences is 5.51% with a doubling time of 12.9 years. (2) In the investigation of the relationship of scientific and economic growth, we focused on UK – one of the few countries with corresponding available (historical) data. The results showed that the scientific growth rate of UK's number of publications (4.97%) is slightly higher than the average worldwide growth rate (4.10%). Furthermore, the results demonstrated that the growth of UK's number of publications is more differentiated (with eight segments) than the worldwide growth (with five segments). The comparison of the British economic and scientific growth rates revealed that the GDP growth rate is lower than the scientific growth rate (3.05% versus 4.97%). Since GDP is not corrected for inflation in this study, results on the comparison of growth rates of science and economy should be interpreted with great care.

In the interpretation of the scientific growth rates that were mostly increasing in the historical development, two interpretations are possible: Either researchers were able to publish more publications in the same time or the increased publication counts can be traced back to an increase in the number of researchers. The study by Fanelli and Larivière (2016) targeted this question. Their results pointed to the second interpretation being more plausible. Fanelli and Larivière (2016) analyzed "individual publication profiles of over 40,000 scientists whose first recorded paper appeared in the Web of Science database between the years 1900 and 1998, and who published two or more papers within the first fifteen years of activity – an 'early-career' phase in which pressures to publish are believed to be high. As expected, the total number of papers published by scientists has increased, particularly in recent decades. However, the average number of collaborators has also increased, and this factor should be taken into account when estimating publication rates. Adjusted for co-authorship, the publication rate of scientists in all disciplines has not increased overall, and has actually mostly declined" (Fanelli & Larivière, 2016).

Two limitations mentioned by Bornmann and Mutz (2015) are still valid for the current study and should be considered in the interpretation of the results:

The first limitation refers to the use of publication counts to measure growth processes. According to Tabah (1999), there are advantages and disadvantages in using these numbers: "although counting publications is simple and relatively straightforward, interpretation of the data can create difficulties that have in the past led to severe criticisms of bibliometric methodology … The main problems concern the least publishable unit (LPU), disciplinary variance, variance in quality of work, and variance in journal quality" (p. 264). The second limitation concerns the interpretation of "growth" as an "increase in numbers". According to Bornmann and Mutz (2015), "it is not clear whether an 'increase in numbers' is directly related to an 'increase of actionable knowledge', for example for reducing needs, extending our knowledge about nature in some lasting way or some other 'higher purposes'" (p. 2221).

Both limitations might be targeted in future studies on growth processes of science. The results of our study show that an exponential growth explains quite well the data and there is different speed in different epochs. However, our study does not target the questions why the growth processes are different and why an exponential growth is present. For example, we show that a regression with five segments have different growth speeds. However, we do not empirically investigate these differences: how can we explain, e.g., that between 1660 and 1793 the growth rate is 3.29%, while between 1793 and 1810 it is 22.78% (Technical Sciences)? Therefore, future studies should try to explore empirically the reasons for different growth processes over time.

This study is based on multi-disciplinary databases only. Future studies that focus on growth processes in various (broad) fields – as we did it in section 3.3 for two broad fields – could use data from mono-disciplinary databases such as Chemical Abstracts (see https://www.cas.org) or Medline (see https://pubmed.ncbi.nlm.nih.gov).

## Acknowledgements

We thank members of the Digital Science team for providing us with feedback on an earlier version of our manuscript.

## Data Availability

The datasets analyzed during the current study are available in the Edmond data repository: https://edmond.mpdl.mpg.de/imeji/collection/D1F8Nf6Sv5aJUKP0 and https://dx.doi.org/10.17617/3.7o. These datasets were derived from the following resources: The Microsoft Academic and Dimensions data used in this paper are from a locally maintained database at the Max Planck Institute for Solid State Research derived from the snapshots provided by Microsoft and Digital Science, respectively. Web of Science and Scopus data were retrieved using the corresponding web-interfaces: https://login.webofknowledge.com and https://www.scopus.com.

## Competing interests

The authors declare no competing interests.

i See http://apps.webofknowledge.com
ii https://www.scopus.com
iii We included "Multidisciplinary" in Life Sciences – following the suggestions by Elsevier – since most of the papers in this category are also assigned to Life Sciences or Health Sciences categories.
iv https://academic.microsoft.com
v https://azure.microsoft.com/
vi See https://app.dimensions.ai/
vii See https://ds.digital-science.com/NoCostAgreement-Collaborators
viii See https://www.abs.gov.au/AUSSTATS/abs@.nsf/Lookup/1297.0Main+Features12008
ix See https://fred.stlouisfed.org/series/NGDPMPUKA

Supplementary Information

Growth rates of modern science: A latent piecewise growth curve approach to model publication numbers from established and new literature databases

Lutz Bornmann, Robin Haunschild, Rüdiger Mutz

Table S1. Results of the mixed effects segmented regression model with four ($M_8$) or five ($M_9$) segments and missing imputation

| Model part | Parameter | All publications ($M_8$) | | | Life Sciences ($M_8$) | | | Physical and Technical Sciences ($M_9$) | | |
|---|---|---|---|---|---|---|---|---|---|---|
| | | Value | Standard error | Doubling time [Year] | Value | Standard error | Doubling time [Year] | Value | Standard error | Double time |
| *Fixed effects* | | | | | | | | | | |
| Intercept | $b_0$ | 4.538 | 1.49* | | 1.017 | 0.845 | | -0.034 | 1.105 | |
| Slopes | | | | | | | | | | |
| Segment 1 | $b_1$ | 0.028 | 0.007* | 24.5 | 0.041* | 0.009 | 17.0 | 0.032 | 0.458 | 21.8 |
| Segment 2 | $b_2$ | 0.055* | 0.003* | 12.6 | 0.079* | 0.026 | 8.8 | 0.226* | 0.050 | 3.1 |
| Segment 3 | $b_3$ | 0.037* | 0.004* | 18.7 | 0.046* | 0.006 | 15.2 | 0.047 | 0.399 | 14.8 |
| Segment 4 | $b_4$ | 0.049541 | 0.003* | 14.0 | 0.047 | 0.022 | 14.8 | -0.0282 | 0.022 | 24.5 |
| Segment 5 | $b_5$ | | | | | | | 0.058* | 0.003 | 12.9 |
| Ending year of segment | | | | | | | | | | |
| Segment 1 | $a_1$ | 1809.1* | 1.23 | | 1808.4* | 0.68 | | 1793.5* | 0.67 | |
| Segment 2 | $a_2$ | 1881.1* | 5.59 | | 1847.6* | 1.35 | | 1809.5* | 1.28 | |
| Segment 3 | $a_3$ | 1952.0* | 2.82 | | 1939.8* | 11.91 | | 1936.2* | 1.95 | |
| Segment 4 | $a_4$ | | | | | | | 1943.3* | 0.83 | |
| *Random effects* | | | | | | | | | | |
| $u_{0j}$ | $\sigma_{u0}$ | 2.802* | 1.026 | | 1.670* | 0.601 | | 2.207* | 0.783 | |
| $u_{1j}$ | $\sigma_{u1}$ | 0.012* | 0.005 | | 0.017* | 0.006 | | 0.696 | 0.729 | |
| | $r_{u1u0}$ | -.97 | 0.028 | | -.958* | 0.046 | | -.885* | 0.109 | |
| $u_{2j}$ | $\sigma_{u2}$ | 0.005* | 0.002 | | 0.050* | 0.018 | | 8.61* | 3.128 | |
| $u_{3j}$ | $\sigma_{u3}$ | 0.009* | 0.003 | | 0.012* | 0.005 | | 36.447 | 86.958 | |
| $u_{4j}/u_{5j}$ | $\sigma_{u4}/\sigma_{u5}$ | 0.005430* | 0.002 | | 0.04* | 0.002 | | 0.645* | 0.233 | |
| $\varepsilon_j$ | $\sigma^2_\varepsilon$ | 0.036* | 0.005 | | 0.0425* | 0.009 | | 0.051* | 0.003 | |

Note. *p<.05

Table S2. Results of the fixed effects segmented regression model ($M_8$) with seven segments for growth domestic product ($M_8$) and eight segments for publication counts ($M_9$) for the UK data

| Model part | Parameter | Gross domestic product | | | Publication counts | | |
|---|---|---|---|---|---|---|---|
| | | Value | Standard error | Doubling time | Value | Standard error | Doubling time |
| Intercept | $b_0$ | 4.245* | 0.024 | | 2.897* | 0.0197 | |
| Slopes | | | | | | | |
| Segment 1 | $b_1$ | 0.008* | 0.001 | 84.6 | 0.075* | 0.002 | 9.7 |
| Segment 2 | $b_2$ | 0.042* | 0.002 | 16.9 | 0.0576* | 0.001 | 12.4 |
| Segment 3 | $b_3$ | 0.005* | 0.002 | 144.8 | 0.0363* | 0.000 | 19.4 |
| Segment 4 | $b_4$ | 0.023* | 0.000 | 30.0 | 0.0519* | 0.003 | 13.7 |
| Segment 5 | $b_5$ | 0.068* | 0.002 | 10.6 | 0.0259* | 0.007 | 27.1 |
| Segment 6 | $b_6$ | 0.135* | 0.005 | 5.5 | 0.0658* | 0.004 | 10.9 |
| Segment 7 | $b_7$ | 0.043* | 0.002 | 16.3 | 0.0830* | 0.001 | 8.7 |
| Segment 8 | $b_8$ | | | | 0.0622* | 0.001 | 11.5 |
| Ending year of segment | | | | | | | |
| Segment 1 | $a_1$ | 1779.8* | 1.333 | | 1805.0* | 1.444 | |
| Segment 2 | $a_2$ | 1810.3* | 1.443 | | 1844.1* | 0.792 | |
| Segment 3 | $a_3$ | 1843.0* | 2.311 | | 1926.7* | 1.648 | |
| Segment 4 | $a_4$ | 1934.8* | 0.968 | | 1939.7* | 1.501 | |
| Segment 5 | $a_5$ | 1968.5* | 0.911 | | 1947.7* | 1.021 | |
| Segment 6 | $a_6$ | 1987.3* | 0.687 | | 1959.4* | 1.789 | |
| Segment 7 | $a_7$ | | | | 1983.1* | 1.117 | |
| $\varepsilon_j$ | | 0.0115 | | | 0.0019 | | |

Note. *$p<.05$

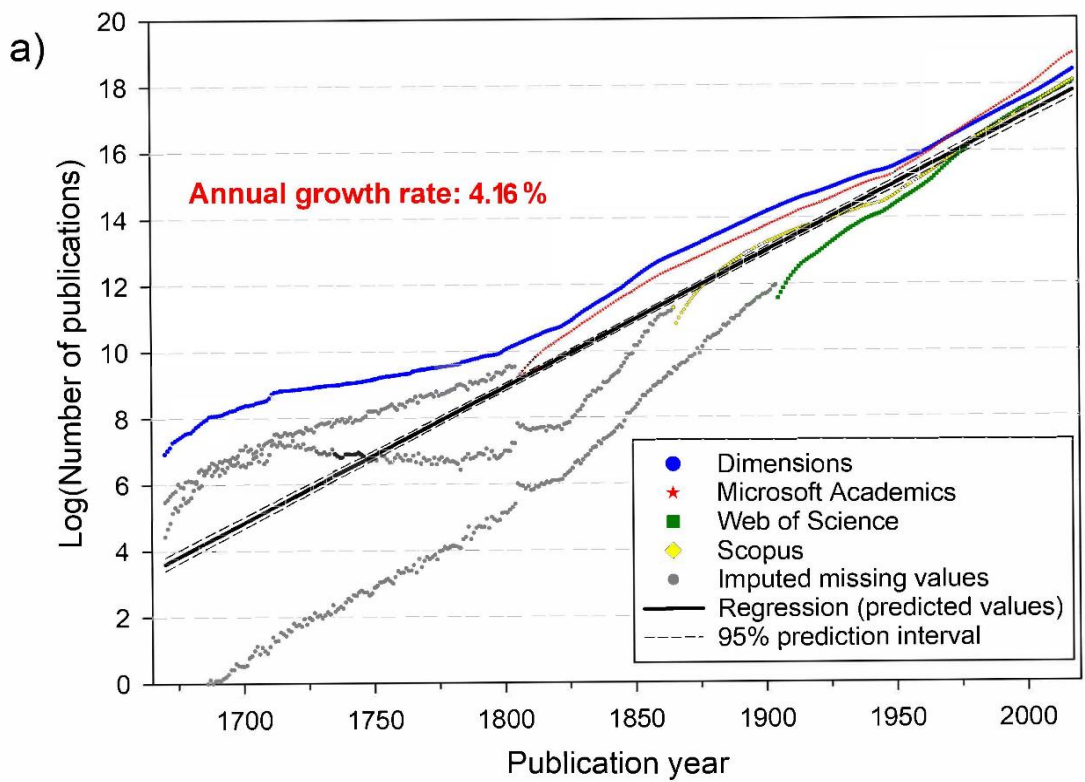

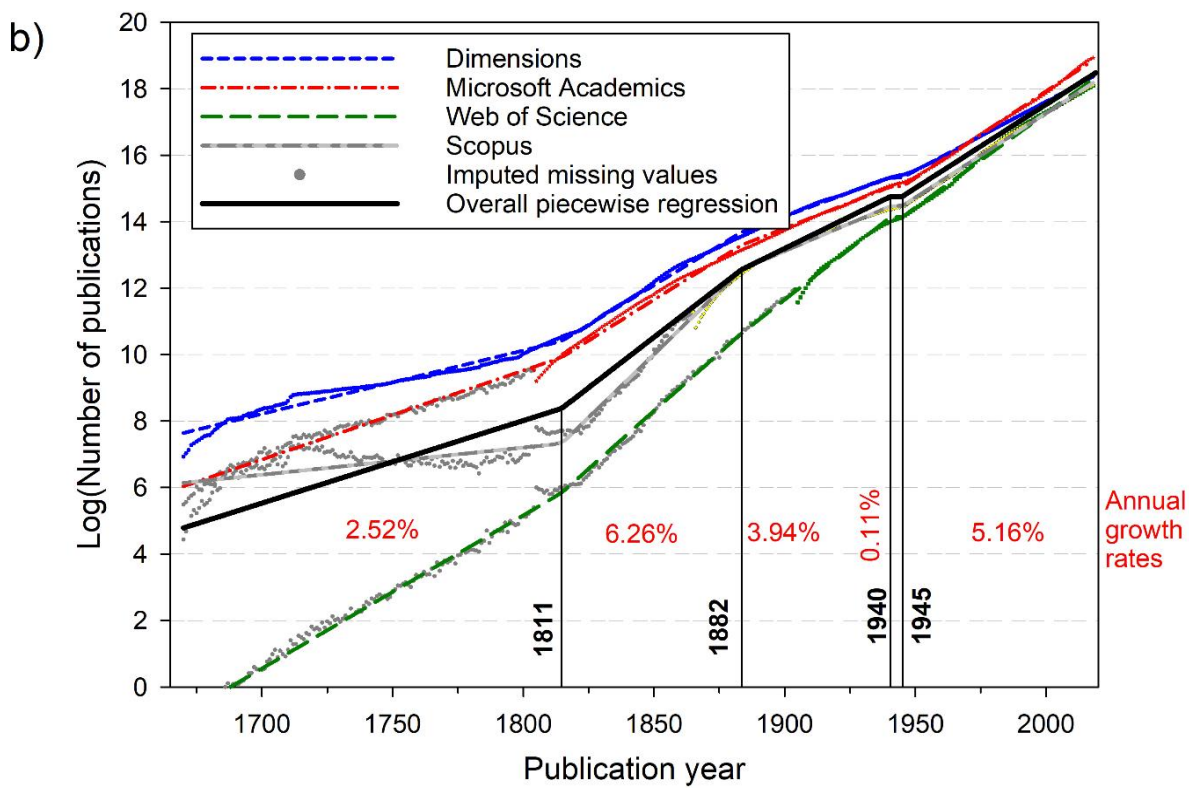

Figure S6. Plots for a) unrestricted growth ($M_1$) and b) segmented unrestricted growth ($M_9$) based on the number of publications from four bibliographic databases with all documents (known and unknown document types)